\documentclass[aps,prd,twocolumn,amsmath,showpacs,superscriptaddress,nofootinbib]{revtex4-1}
\usepackage{graphicx}
\usepackage{amsmath}
\usepackage{longtable}
\usepackage{float}
\usepackage{dcolumn}
\usepackage{bm}
\usepackage{appendix}
\usepackage{multirow}
\usepackage{color}
\usepackage{soul}
\usepackage[normalem]{ulem}

\usepackage{hyperref}

\hypersetup{colorlinks,linkcolor={blue},citecolor={blue},urlcolor={blue}}

\begin{document}

\title{\boldmath Visual tool for assessing tension-resolving models in the $H_0$-$\sigma_8$ plane}

\author{Igor de O. C. Pedreira}
\email{igorpedreira@id.uff.br}
\affiliation{Instituto de F\'{i}sica, Universidade Federal Fluminense,
Avenida General Milton Tavares de Souza s/n, Gragoat\'{a}, 24210-346 Niter\'{o}i, Rio de Janeiro, Brazil}

\author{Micol Benetti}
\email{micol.benetti@unina.it}
\affiliation{Scuola Superiore Meridionale (SSM), Universit\`{a} di Napoli ``Federico II'', Largo San Marcellino 10, I-80138 Napoli, Italy}
\affiliation{Istituto Nazionale di Fisica Nucleare (INFN), Sezione di Napoli, Via Cinthia 9, I-80126 Napoli, Italy}

\author{Elisa G. M. Ferreira}
\email{elisa.ferreira@ipmu.jp}
\affiliation{Kavli Institute for the Physics and Mathematics of the Universe (WPI),
UTIAS, The University of Tokyo, Chiba 277-8583, Japan}
\affiliation{Instituto de F\'isica, Universidade de S\~ao Paulo,C.P. 66318, CEP: 05315-970, S\~ao Paulo, Brazil}
\affiliation{Center for Data-Driven Discovery, Kavli IPMU (WPI),
UTIAS, The University of Tokyo, Kashiwa, 277-8583, Japan}

\author{Leila L. Graef}
\email{leilagraef@id.uff.br}
\affiliation{Instituto de F\'{i}sica, Universidade Federal Fluminense,
Avenida General Milton Tavares de Souza s/n, Gragoat\'{a}, 24210-346 Niter\'{o}i, Rio de Janeiro, Brazil}

\author{Laura Herold}
\email{lherold@jhu.edu}
\affiliation{Max-Planck-Institut f\"ur Astrophysik, Karl-Schwarzschild-Str.\ 1, 85748 Garching, Germany}
\affiliation{Department of Physics and Astronomy, Johns Hopkins University,
3400 North Charles Street, Baltimore, Maryland 21218, USA}

\begin{abstract}
\noindent Beyond-$\Lambda$CDM models have been proposed to address various shortcomings of the standard cosmological model, such as the ``Hubble tension.'' These models often have an impact on the discrepancy in the amplitude of matter clustering, the ``$\sigma_8$-tension.'' To explore the interplay between the two tensions, we suggest a simple method to visualize the relation between the two parameters: $H_0$ and $\sigma_8$. For a given extension of the $\Lambda$CDM model and dataset, we plot the relation between $H_0$ and $\sigma_8$ for different amplitudes of the beyond-$\Lambda$CDM physics. In this work, we use this visualization method to illustrate the trend of selected cosmological models, including nonminimal Higgs-like inflation, early dark energy, a varying effective electron mass, an extra number of relativistic species and modified dark energy models. Although already studied in the literature, some of these models have not been analyzed in view of the two joint tensions. We stress that the method used here could be a useful diagnostic tool to illustrate the behavior of complex cosmological models with many parameters in the context of the $H_0$ and $\sigma_8$ tensions.
\end{abstract}

\date{\today}

\maketitle

\section{Introduction}
\label{sec:intro}
With the improvement of both theoretical modeling and observations of our Universe, our understanding of cosmology has undergone a revolution in the past decades. 
The standard cosmological model ($\Lambda$CDM) is a parametrization which is capable of explaining a large part of the evolution of the Universe, its composition, and the structures we see today. This model is extremely successful in 
describing observations~\cite{WMAP:2012nax,Planck:2018vyg,Cyburt:2004yc,Anderson:2012sa,SupernovaSearchTeam:1998fmf, Riess:2019cxk, Riess:2021jrx, Scolnic:2021amr, Brout:2022vxf}, while having most of its six parameters measured with subpercent precision.
However, with the increase in measurement precision, parameter discrepancies have appeared, which have become statistically significant with the latest data analyses. The most significant of these discrepancies is the ``Hubble tension'' or ``$H_0$ tension.'' Measurements of the present-day expansion rate of the Universe, the Hubble constant, $H_0$, obtained via indirect measurements, which depend on the assumption of a cosmological model, yield systematically lower values of $H_0$ than direct measurements, which do not or weakly depend on the assumption of a cosmological model. The most significant tension is seen between the (indirect) inference of $H_0$ from cosmic microwave background (CMB) data of the \textit{Planck} mission assuming the $\Lambda$CDM model~\cite{Planck:2018vyg}, $H_0 = 67.4\pm0.5\, \mathrm{km/s/Mpc}$, and the (direct) local inference from Cepheid-calibrated type Ia supernovae of the SH$0$ES project~\cite{Riess:2021jrx}, $H_0 = 73.0\pm1.0 \, \mathrm{km/s/Mpc}$. Considering these measurements, the statistical significance of the tension is currently at $5 \sigma$. A consistently higher value of $H_0$ is also present in other local measurements of $H_0$ and the tension can vary from $4\sigma$-$6\sigma$~\cite{DiValentino:2021izs}, although studies using the tip of the red-giant branch as calibrators instead of Cepheids~\cite{Freedman:2019jwv, Freedman:2021ahq, Scolnic:2023mrv, Anand:2021sum, madore2023population}, found a value of $H_0$ between the direct and indirect measurements.

Another discrepancy has been found in current measurements sensitive to the amount of matter clustering in our Universe, represented by the clustering parameter $\sigma_8$ or the related parameter $S_8$, where $S_8= \sqrt{\Omega_m/0.3}\,\sigma_8$ with $\Omega_m$ being the matter energy density. This “$\sigma_8$ tension’’ 
(see \cite{Abdalla:2022yfr} for a recent review)
describes the discrepancy between the higher inferred value of $\sigma_8$ or $S_8$ from CMB data from \textit{Planck} ($\sigma_8 = 0.8111 \pm 0.0060$ and $S_8 = 0.832 \pm 0.013$ for TT,TE,EE+lowE+lensing~\cite{Planck:2018vyg}) assuming the $\Lambda$CDM model, and the lower values obtained in low-redshift probes such as weak gravitational lensing and galaxy clustering. Depending on the data considered this tension ranges from $2-4 \sigma$, with galaxy shear presenting the largest discrepancy with CMB. 
For the Kilo-Degree Survey (KiDS-1000), this discrepancy is at the level of $ \sim 2.4-2.7\sigma$\footnote{The level of significance depends on the details of the analysis, and reaches $3.1\sigma$ in some cases.} where cosmic shear yields $\sigma_8 = 0.838_{-0.141}^{+0.140}$ ($S_8 = 0.759^{+0.024}_{-0.021}$) \cite{KiDS:2020suj}, and the $3 \times 2 $pt analysis that combines shear-shear, galaxy-galaxy lensing and galaxy-galaxy two-point statistics, yields $\sigma_8 = 0.7600^{+0.025}_{-0.020}$ ($ S_8 = 0.766^{+0.020}_{-0.014}$) \cite{Heymans:2020gsg}.
For the Dark Energy Survey (DES) Year 3 (Y3), the discrepancy is on the $\sim 2.3-2.6\sigma$ level \cite{DES:2021bvc,DES:2021wwk}. The Hyper Suprime-Cam (HSC) collaboration reports a compatible statistical significance with CMB to the previous clustering probes ($\sim 2\sigma$) in its year 3 analysis~\cite{Dalal:2023olq,Li:2023tui}, and the $3 \times 2 $pt analysis indicates no significant tension with Planck's $S_8$ value~\cite{More:2023knf,Miyatake:2023njf,Sugiyama:2023fzm}.
Recently, a combined analysis of DES and KiDS-1000 cosmic-shear data \cite{Kilo-DegreeSurvey:2023gfr} obtained $\sigma_8 = 0.825^{+0.067}_{-0.073}$ ($S_8 = 0.790^{+0.018}_{-0.014}$), which is consistent with the \textit{Planck} measurements at $1.7\sigma$, and with other clustering probes, like HSC-Y3.
Besides its low significance, the $\sigma_8$ tension has been consistently present in the results from independent photometric surveys like KiDS, DES, and HSC with similar levels of significance, leading to a persistent interest in the $\sigma_8$ tension \cite{Benisty:2020kdt, Nunes:2021ipq}. On the other hand, the Baryon Oscillation Spectroscopic Survey (BOSS) and extended BOSS surveys using spectroscopic data find values of $\sigma_8$ consistent with both \textit{Planck} and the other clustering surveys. 

The origin of these tensions is still unknown. While they could be a result of unknown measurement systematics, they could also hint at new physics beyond the $\Lambda$CDM model \cite{Kitching:2016hvn, DiValentino:2020zio, DiValentino:2020vvd, Riess:2023bfx, Freedman:2023jcz, Vagnozzi:2023nrq}.
Many models were proposed in the literature with the goal of solving one or even both of these tensions. 
For the Hubble tension, these models can be classified as early time solutions when the new physics is added pre-recombination, and late time solutions for post-recombination extensions. The early time solutions aim to decrease the physical size of the sound horizon at last scattering, which leads to an increase in $H_0$. This can be done in different ways, for example, by increasing $H(z)$ before recombination with additional components, by changing the redshift of last scattering or the redshift of matter-radiation equality by adding new physics in the pre-recombination era \cite{Benetti:2019lxu, Borges:2023xwx, Benetti:2021div, Salzano:2021zxk, Benevento:2022cql, Kable:2023bsg}, or by changing the sound speed of the baryon/photon plasma (for a review see \cite{Schoneberg:2021qvd,Knox:2019rjx}). One can also consider even earlier modifications of the physics during the inflationary era that could lead to a higher $H_0$ \cite{Rodrigues:2023kiz, Rodrigues:2021txa, Rodrigues:2020dod}.
Moreover, late time solutions to the $H_0$ tension aim to increase the current rate of expansion directly \cite{DiValentino:2021izs, Abdalla:2022yfr, Salzano:2021zxk, Montani:2023xpd, Adil:2021zxp, Frion:2023xwq, Bernui:2023byc}. 
On the other hand, most of the models proposed to solve the $\sigma_8$ tension are either based on decreasing the predicted value of $\Omega_m$ or on a late time suppression of the linear matter power spectrum \cite{Anchordoqui:2021gji}.\footnote{For recent late time analysis in the context of both tensions, see, for instance, Refs.~\cite{Banerjee:2020xcn, Heisenberg:2022lob, Heisenberg:2022gqk, DAgostino:2023cgx, Gangopadhyay:2023nli}.}

However, the majority of models proposed to solve one of the tensions exhibit a positive correlation between $H_0$ and $\sigma_8$, where an increase in $H_0$ leads to an increase in $\sigma_8$, and vice versa \cite{DiValentino:2021izs,deSa:2022hsh, Jedamzik:2020zmd}, leading to a relaxation of one of the tensions while the other one is worsened. Furthermore, many works do not analyze both tensions simultaneously, possibly relaxing one without giving information about the other.There are, nevertheless, some classes of models that attempt to alleviate both tensions. Among them, we can mention some classes of late interacting dark-energy-dark-matter (IDE) models \cite{Benetti:2019lxu, Borges:2023xwx, Benetti:2021div, Salzano:2021zxk}, some new early dark energy models \cite{Cruz:2023lmn}, modification in the standard model at early times \cite{Rodrigues:2020dod,Rodrigues:2021txa,Rodrigues:2023kiz}, models with a late time change to the equation-of-state parameter of dark matter \(\omega_\mathrm{cdm}\) \cite{Naidoo:2022rda}, among other proposals (see also the model of Ref.~\cite{Basilakos:2023kvk} for instance). However, such models still deserve further investigation in order to be confirmed as viable solutions to the tensions.

Given the large number of models in the literature proposed to address the tensions, in this article, we suggest a simple method to visualize proposed tension-resolving models in the $H_0-\sigma_8$ plane to assess the tendency of the models regarding the $H_0$, $\sigma_8$, or both tensions. 
Although the relation between $H_0$ and $\sigma_8$ is complicated and depends on all parameters of the model and the chosen dataset, one can find an empirical relation between the resulting value of $H_0$ and $\sigma_8$ from parameter inference runs where one or more of the parameters that represent the $\Lambda$CDM extension are fixed. By fixing these extra parameters to different values, we can plot the resulting values of $H_0$ and $\sigma_8$. The result is an approximate line in the $H_0-\sigma_8$ plane. 
This method was inspired by the work in Ref.~\cite{Vagnozzi:2019ezj} and was also presented in~\citep{deSa:2022hsh}. Taking at face value both tensions, one can use this representation as a diagnostic tool to assess the possibility of beyond $\Lambda$CDM models to solve both the $H_0$ and the $S_8$ tensions. 
We use this representation to test the correlation of interesting prospective tension-solving models. In the results section we show the tendency of these models in solving one or both tensions. Among the models chosen, for 
the HLI and varying electron mass (and curvature), this tendency is shown explicitly for the first time here, showing the usefulness of this method.

The goal of this paper is not to give a definite answer as to which models solve the tensions, as there are many comprehensive analyses that test various models in the context of the $H_0$, $\sigma_8$, or both tensions, including some of the models chosen here \cite{Schoneberg:2021qvd,Knox:2019rjx}. Given the complexity of these models and the dependence of $H_0$ and $\sigma_8$ on each other and on other cosmological parameters, this method has the goal to provide an intuitive illustration of how the values of $H_0$ and $\sigma_8$ vary with the extra beyond-$\Lambda$CDM parameters and diagnose the tendency of these models with respect to the $H_0$ and $\sigma_8$ tensions.
The paper is organized as follows: In Sec.~\ref{sec:Method}, we describe the methodology of the analysis, presenting the construction of the $H_0-\sigma_8$ plane. In Sec.~\ref{sec:Theory}, we introduce the models we selected to test the method. We present the results in Sec.~\ref{sec:Results} and conclude in Sec.~\ref{sec:Conclusions}.

\section{Methodology}
\label{sec:Method}


In this section, we introduce the methodology for constructing the $H_0$-$\sigma_8$ plane and give details about the analysis choices and datasets used in the statistical analysis. 

\subsection{Constructing the $H_0$-$\sigma_8$ plane}

The parameters $H_0$ and $\sigma_8$ have a significant degeneracy with the cosmological parameters and each other, respectively. This degeneracy depends both on the dataset and on the cosmological model that is considered.
Therefore, it is nontrivial to understand how introducing new physics beyond the $\Lambda$CDM model will affect the relationship between $H_0$ and $\sigma_8$. 
In this work, we suggest a method to visualize the relation between these parameters in order to explore the prediction of beyond-$\Lambda$CDM models with respect to the $H_0$ and $\sigma_8$ tensions.

To construct the $H_0$-$\sigma_8$ relation, we perform a statistical analysis (detailed in Sec.~\ref{sec:inference_and_data}) under the selected datasets while keeping the extra parameter(s) of the chosen beyond-$\Lambda$CDM model fixed to different values. The fixed parameter is chosen such that it characterizes the extension of $\Lambda$CDM, and its values are quoted for each model in Sec.~\ref{sec:Theory}. For each value of the fixed parameter, we infer the best fit or mean values of $H_{0}$ and $\sigma_{8}$, which gives a point in the $H_0$-$\sigma_8$ plane. Repeating the inference for different values of the fixed parameter yields a $H_0$-$\sigma_8$ line in the plane.
The $H_0$-$\sigma_8$ line provides an intuitive illustration of the behavior of the given beyond-$\Lambda$CDM model with respect to the two tensions and can be used to compare to the measured values of, e.g., $H_0$ from the SH$0$ES collaboration \cite{Riess:2021jrx}, $\sigma_8$ from KiDS-1000 3x2pt analysis~\citep{Heymans:2020gsg}, and to $H_0$ and $\sigma_8$ from the \textit{Planck} collaboration \cite{Planck:2019nip}. 
Equivalently, one could construct the $H_0$-$S_8$ plane, using the $S_8$ parameter obtained in the statistical analysis, which contains similar information about the behavior of the beyond-$\Lambda$CDM model with respect to the tensions\footnote{In our case, the $H_0$-$S_8$ plane was slightly noisier than using $\sigma_8$, so we opted for the latter.}.

It is important for this method to choose the fixed extra parameter from a large range of values. This allows 
to detect parts of the parameter space that might have a different correlation than 
other parts of the parameter space (as seen in \cite{Kreisch:2019yzn} for example). 
One could for example imagine an island, where the values of $H_0$ and $\sigma_8$ do not fall into the line constructed with the other points. These regions in parameter space might be interesting to study but might represent parts of the parameter space that do not fit the data well. 

\subsection{Statistical inference and datasets}
\label{sec:inference_and_data}

We perform our analyses using the Boltzmann-solver codes \textit{Code for Anisotropies in the Microwave Background} ({\sc CAMB}) \cite{Lewis:1999bs} and \textit{Cosmic Linear Anisotropy Solving System} (CLASS) \cite{Blas:2011rf}. We 
vary the six free parameters of the $\Lambda$CDM model, i.e. the baryon density ($\Omega_bh^2$), the cold dark matter density ($\Omega_\mathrm{cdm}h^2$), the ratio between the sound horizon and the angular diameter distance at decoupling ($\theta$), the optical depth ($\tau$), the primordial scalar amplitude ($A_s$), and the primordial spectral index ($n_s$). We follow the convention of the \textit{Planck} collaboration~\cite{Planck:2018vyg} and assume two massless and one massive neutrino with $m_\nu = 0.06$ eV (except for the models with varying neutrino masses). In addition, the beyond-$\Lambda$CDM model introduces a number of extra parameter(s). To construct the $H_0-\sigma_8$ plane, we fix one (or two) of these extra parameters at several values of interest and perform a statistical inference.

We infer the posterior mean values of $H_0$ and $\sigma_8$ by conducting a Markov chain Monte Carlo (MCMC) analysis, using {\sc CosmoMC}~\cite{Lewis:2002ah} and \texttt{MontePython} \cite{Audren:2012wb, Brinckmann:2018cvx}. To test the chain convergence, we use the Gelman-Rubin convergence criterion, requiring $R-1 < 0.05$. For all models introduced in Sec.~\ref{sec:Theory}, we perform an MCMC analysis keeping the extra parameter fixed to different values. The only exceptions are the early dark energy (EDE) and EDE+$\nu$ models, where we infer the maximum-likelihood (or best fit) values of $H_0$ and $\sigma_8$ using a minimization algorithm keeping the extra parameter fixed to different values (similar to a frequentist profile likelihood analysis). We use the minimization for these models since we want to avoid an influence of prior volume effects
\footnote{Prior volume effects, also known as marginalization effects, are a consequence of the marginalization in an MCMC analysis, which can appear if the data do not constrain all parameters of the model well. This can lead to an up-weighting of regions with larger prior volume regardless of the goodness of fit in the respective region.} \cite{Murgia:2020ryi, Niedermann:2020dwg, Smith:2020rxx, Herold:2021ksg,Reeves:2022aoi, Gomez-Valent:2022hkb}. 
In the statistical inference, we use combinations of the following datasets:
\begin{itemize}
\item \textit{Planck}: the CMB 
data from \textit{Planck} 2018 \cite{Planck:2019nip} 
considering both temperature and temperature-polarization cross-correlation likelihoods with temperature and polarization EE likelihood at $\ell \leq 29$, as well as the CMB lensing reconstruction power spectrum ~\cite{Planck:2019nip,Planck:2018lbu}, referred to as `\textit{Planck}'.
\item BAO: post-reconstruction baryon acoustic oscillation (BAO) data from the BOSS Data Release 12~\cite{Alam:2016hwk}, from 6dF Galaxy Survey (6dFGS) ~\cite{Beutler:2011hx}, and from SDSS DR7 Main Galaxy Sample (SDSS-MGS) galaxies~\cite{Ross:2014qpa}, referred to as ``\textit{reconstructed BAO}.'' We also use the full shape of the galaxy power-spectrum multipoles based on the effective field theory of large scale structure~\cite{Baumann:2010tm, Carrasco:2012cv, Ivanov:2019pdj, Chudaykin:2020aoj, Philcox:2020vvt} with the same settings as in~\cite{Herold:2022iib}, referred to as ``\textit{full-shape BOSS}.'' 

\item Pantheon: the Pantheon sample of Supernovae (SNe) Type Ia~\cite{Scolnic:2017caz}, which includes 1048 SNe in the redshift range $z = 0.01-2.26$, referred to as ``\textit{Pantheon}.''
\end{itemize}

\noindent
We use the joint datasets of \textit{Planck}, reconstructed BAO and Pantheon data for our analysis. The only exception are the EDE and EDE+$\nu$ models, where we use \textit{Planck} and full-shape BOSS data to ensure consistency with the analysis performed in the previous work \cite{Reeves:2022aoi, Herold:2021ksg}. We tested that including the post-reconstruction BAO data from BOSS did not significantly change the results. 
We do not expect that this slight change in the dataset impacts our conclusions. 

\section{Selected Models}
\label{sec:Theory}

For our analysis, we select several extensions of the $\Lambda$CDM model that have been proposed as possible solutions to the $H_0$ tension or of both $H_0$ and $\sigma_{8}$ tensions. These models can be broadly classified into two classes: early time and late time solutions, relating to models that modify the $\Lambda$CDM model at the early Universe and large scale, respectively. 
Among the early time solutions there are models that modify the inflationary physics like the nonminimal Higgs-like inflation in the strong coupling regime model \cite{Rodrigues:2023kiz, Rodrigues:2021txa};
or models that change pre-recombination physics like the early dark energy model \cite{Kamionkowski:2014zda, Karwal:2016vyq, Caldwell:2017chz, Smith:2019ihp, Pettorino:2013ia, Poulin:2018cxd};
a model that varies the fundamental constant of the electron mass \cite{Schoneberg:2021qvd, Hart:2019dxi, Hart:2017ndk, Sekiguchi:2020teg};
or models that have extra numbers of relativistic species \cite{DiValentino:2020zio, DiValentino:2016hlg,Bernal:2016gxb,Benetti:2017gvm,Benetti:2017juy,Mortsell:2018mfj,DEramo:2018vss,Guo:2018ans,Kreisch:2019yzn,Vagnozzi:2019ezj,Ballesteros:2020sik,RoyChoudhury:2020dmd,Brinckmann:2020bcn,Seto:2021xua,DiValentino:2021izs, DiValentino:2015wba, Benetti:2021uea, deSa:2022hsh,Archidiacono:2013fha,Alcaniz:2022oow}. 
Alternatively, there are late time solutions, like modified dark energy, which have an equation of state $w \neq -1$ \cite{DiValentino:2016hlg, DiValentino:2020zio, Bargiacchi:2021hdp, Gonzalez:2021ojp, deSa:2022hsh, Yang:2018qmz, Yang:2018prh, DiValentino:2019dzu, Vagnozzi:2019ezj, DiValentino:2020naf, Keeley:2019esp, Joudaki:2016kym, Zhang:2018air, Visinelli:2019qqu, DiValentino:2019jae, Ludwick:2017tox}. 
There is extensive discussion in the literature about whether early or late time solutions are more effective in resolving the Hubble tension, with most of these works concluding that early time solutions perform better in increasing the value of $H_0$ \cite{Schoneberg:2021qvd, Knox:2019rjx}, and others conclude that a combination of early and late time solutions is necessary \cite{Vagnozzi:2023nrq}.

The models adopted here were chosen to provide a heterogeneous sample of different behaviors with respect to the tensions but are by no means complete. This allows us to illustrate the method and diagnose the models' tendency to alleviate both tensions. The inferred values of $H_0$ within the models present different levels of agreement or tension with the SH$0$ES measurement as is presented in the respective literature of these models. Below we describe the models under consideration and the mechanisms that allow them to address the tensions. We highlight the extra parameters of these models beyond the six $\Lambda$CDM ones $\left\{ \omega_b, \omega_{\mathrm{cdm}}, \theta_s, n_s, A_s, \tau \right\}$, and which of them are fixed in the analysis.

\begin{itemize}
\item{\textbf{nonminimal Higgs-Like Inflation (HLI):}}
We examine an inflationary model where a nonminimal coupling, $\xi$, between the inflaton field and the Ricci scalar is considered as well as radiative corrections at one loop order \cite{Rodrigues:2023kiz, Rodrigues:2021txa, Rodrigues:2020dod}. Those inflationary models incorporate a generic seesaw extension (types I and II) of the Standard Model of particle physics. When assuming a strong coupling regime, it has been noted that the usual $H_0$-$\sigma_8$ correlation is 
broken if considering an inflationary phase with duration $N_\mathrm{e-folds}=\mathcal{N} >54.5$ \cite{Rodrigues:2023kiz}. 

Here, we 
consider the strong coupling regime and fix $\xi=100$, while fixing the duration to different values $\mathcal{N} = \left\{ 55, 56, 58, 60\right\}$. This is a five dimension model since the free parameters of the theory are the $\Lambda$CDM parameters $\left\{ \omega_b, \omega_\mathrm{cdm}, \theta_s, \tau \right\}$, while the primordial spectrum is not expressed in terms of {$A_s$, $n_s$} but directly related to the inflationary potential, whose free parameter is the deviation from the tree level potential, $a'$. In the plots and where it is convenient, we call this model ``HLI'' for simplicity. 
In Table \ref{tab:HLI}, we report the values obtained in Ref.~\cite{Rodrigues:2023kiz} for the relevant parameters, which are going to be used in our analysis.

\item{\textbf{Early Dark Energy (EDE):} }
The EDE models \cite{Pettorino:2013ia, Kamionkowski:2014zda, Karwal:2016vyq, Caldwell:2017chz, Poulin:2018cxd,Smith:2019ihp, Kamionkowski:2022pkx, Poulin:2023lkg, McDonough:2023qcu} are a class of models that consider a new component for the energy density of the Universe. This component behaves like a cosmological constant before recombination but then decays away faster than matter, leading to an increase in the expansion rate just prior to recombination. This reduces the physical size of the sound horizon at last scattering, allowing for a higher $H_{0}$ from CMB data without conflicting late time constraints from BAO and Hubble-flow SNe Ia.

In this work, we will consider the model discussed in \cite{Poulin:2018cxd,Smith:2019ihp}, which assumes a pseudoscalar field with a potential that describes the above characteristics. This model has three extra parameters $\left\{ f_{\mathrm{EDE}}, z_c, \theta_i \right\}$, where $f_{\mathrm{EDE}}$ is the maximum fraction of EDE at a critical redshift $z_c$, and $\theta_i=\phi_i/f$ is the initial value of the dimensionless field. For this analysis, we set $f_\mathrm{EDE}$ to 
$f_{\mathrm{EDE}} = \left\{0.00, 0.05, 0.10, 0.15\right\}$. In the following, we refer to this model as ``EDE.'' In Table \ref{tab:EDE-EDE+nu}, we report the values of the relevant parameters obtained in Refs.~\cite{Herold:2021ksg, Herold:2022iib}, which are going to be used in the analysis performed here.

\item{\textbf{EDE + massive neutrinos:}}
We consider also an extension of the EDE model, where the total neutrino mass $m_\nu$ is a free parameter. Allowing $m_\nu$ to be a free parameter in the statistical analysis was proposed as a possible solution to the larger clustering amplitude present in EDE models \cite{Poulin:2018zxs, Reeves:2022aoi}. Due to their free-streaming nature, massive neutrinos suppress small-scale power, lowering the $\sigma_8$ parameter \cite{Brandbyge:2010ge}.

This model has four extra parameters $\left\{ f_{\mathrm{EDE}}, z_c, \theta_i, m_\nu \right\}$, where $m_\nu$ is fixed to different values in the analysis: $m_{\nu} = \left\{ 0.06, 0.09, 0.15, 0.24 \right\}$. We call this model ``EDE $+ \nu$'' for simplicity. In Table \ref{tab:EDE-EDE+nu}, we report the values for the relevant parameters obtained in Ref.~\cite{Reeves:2022aoi}, which are going to be used in the analysis performed here.

\item{\textbf{Varying Effective Electron Mass:}}
A variation of the fundamental properties of the hydrogen/helium atom, such as the electron mass, is one effective way to shift the time of recombination in the early Universe. Shifting the energy gap between successive excitation levels implies changing the temperature at which the photodissociation of hydrogen/helium becomes inefficient. Therefore, there is a strong degeneracy between variations of these fundamental parameters and the redshift of recombination \cite{Sekiguchi:2020teg}.

Here, we follow the approach of \cite{Schoneberg:2021qvd} and \cite{Hart:2019dxi} in allowing for a spatially uniform time-independent variation in the electron mass. 
In these models, the characteristic parameter describes the variation in the electron mass. It is defined by
\begin{equation}
\delta m_\mathrm{e} \equiv \frac{m_\mathrm{e,early}}{m_\mathrm{e,late}},
\end{equation}
where $m_\mathrm{e,early}$ and $m_\mathrm{e,late}$ are the values of $m_\mathrm{e}$ inferred, respectively, from early and late times (observed locally). We consider the variation of $m_\mathrm{e,early}(z)$ as being redshift independent over recombination. 
This model has one extra parameter $\left\{ \delta m_\mathrm{e} \right\}$, which we fix to different values in the analysis: $\delta m_\mathrm{e} = \left\{0.96, 1.00, 1.06\right\}$. 

We call this model ``Varying $m_\mathrm{e}$'' for simplicity. In Table \ref{tab:me_me+k}, we report the values for the parameters that are going to be used in the analysis performed here. 

\item{\textbf{Extra Number of Effective Relativistic Species:}}
These models consider extra contributions to the relativistic components of the Universe. This extra contribution could come from extra relativistic neutrinos from particle physics, from a significant stochastic background of gravitational waves, or from the decay products of other components in our Universe~\cite{deSa:2022hsh, DiValentino:2016hlg, Bernal:2016gxb, Benetti:2017gvm, Benetti:2017juy, Mortsell:2018mfj, DEramo:2018vss, Guo:2018ans, Kreisch:2019yzn, Vagnozzi:2019ezj, Ballesteros:2020sik, DiValentino:2020zio, RoyChoudhury:2020dmd, Brinckmann:2020bcn, Seto:2021xua, DiValentino:2021izs, DiValentino:2015wba, Benetti:2021uea, Archidiacono:2013fha, Alcaniz:2022oow}. 

This model has one extra parameter $\left\{N_\mathrm{eff}\right\}$, which we fix to different values in the analysis: $N_\mathrm{eff} = \left\{3.046, 3.15, 3.55, 3.95\right\}$.
We refer to this model as `Varying $N_\mathrm{eff}$'.
In Table \ref{tab:Neff}, we report the values obtained in Ref.~\cite{deSa:2022hsh} for the relevant parameters, which are going to be used in the analysis performed here. 

\item{\textbf{Varying Effective Electron Mass and Curvature:}} 
The presence of nonzero curvature, $\Omega_{k}$, in combination with a varying electron mass seems to succeed in reducing the Hubble tension \cite{Schoneberg:2021qvd,Sekiguchi:2020teg}. This can be understood since in the original model, in addition to shifting the contribution to $r_{s}$, it also simultaneously shifts the corresponding angular diameter distance, $D_{A}$, for the CMB, but not for the BAO and other late time probes. Both these changes can be absorbed with the usual $\Lambda$CDM parameters, however, it is impossible to absorb both simultaneously. Nevertheless, the angular diameter distances to the BAO and CMB are impacted in distinct ways by an increase in $\Omega_{k}$. Therefore, in this two-parameter extension $\left\{\delta m_\mathrm{e} + \Omega_{k}\right\}$ of $\Lambda$CDM, it is possible to preserve both angular scales of the BAO and the CMB under a variation of the redshift of recombination.
This model and the previous one are interesting since they were considered to be the most successful models to relax the tension in $H_0$ according to the criteria adopted in the work of Ref.~\cite{Schoneberg:2021qvd}. 

We refer to this model as ``Varying $\delta m_\mathrm{e} + \Omega_{k}$'', and in our analysis, we consider only the case with its two extra parameters fixed, $\delta m_\mathrm{e} = 1.06$, $\Omega_{k}$={0.01}. This allows us to test whether adding a positive $\Omega_k$ changes the direction of the flat ``varying $m_\mathrm{e}$'' model in the $H_0$-$\sigma_8$ plane. The results are reported in Table \ref{tab:me_me+k}.

\item{\textbf{Modified Dark Energy (Phantom and Quintessence):}}
This class of models considers the possibility of a dark energy component with an energy density evolving in time, in contrast to the cosmological constant behavior. In these scenarios, the equation-of-state parameter, $w$, of the dark energy can assume values different from the canonical $w=-1$. There are models in the literature that can describe this behavior in a phenomenological way and in concordance with several observations \cite{Bargiacchi:2021hdp, Joudaki:2016kym, Zhang:2018air, Visinelli:2019qqu, DiValentino:2019jae, Keeley:2019esp,Ludwick:2017tox}. 

This model has one extra parameter $\left\{w\right\}$, which we will 
fix to different values: $w = \left\{-0.7, -0.8, -0.9, -1.0, -1.1, -1.2, -1.3\right\}$.
In the plots and where it is convenient, we call this model ``phantom DE'' when $w < -1$, and ``quintessence'' when $w > -1$. In Table \ref{tab:w}, we report the values of the parameters that are going to be used in our analysis. Those include the values obtained in Ref.~\cite{deSa:2022hsh} for the phantom regime, in addition to the values obtained here with $w > -1$ for the quintessence regime.
\end{itemize}

\section{Results and Discussions}
\label{sec:Results}

In Sec.~\ref{sec:result1}, we present the $H_0-\sigma_8$ plane for the models introduced above, and in Sec.~\ref{sec:chi2}, we describe how the goodness of fit to the data can be included as an additional layer of information.

\subsection{Visualizing models in the $H_0$-$\sigma_8$ plane}
\label{sec:result1}
 
\begin{figure*} [t!]
  \includegraphics[width=0.9\textwidth]{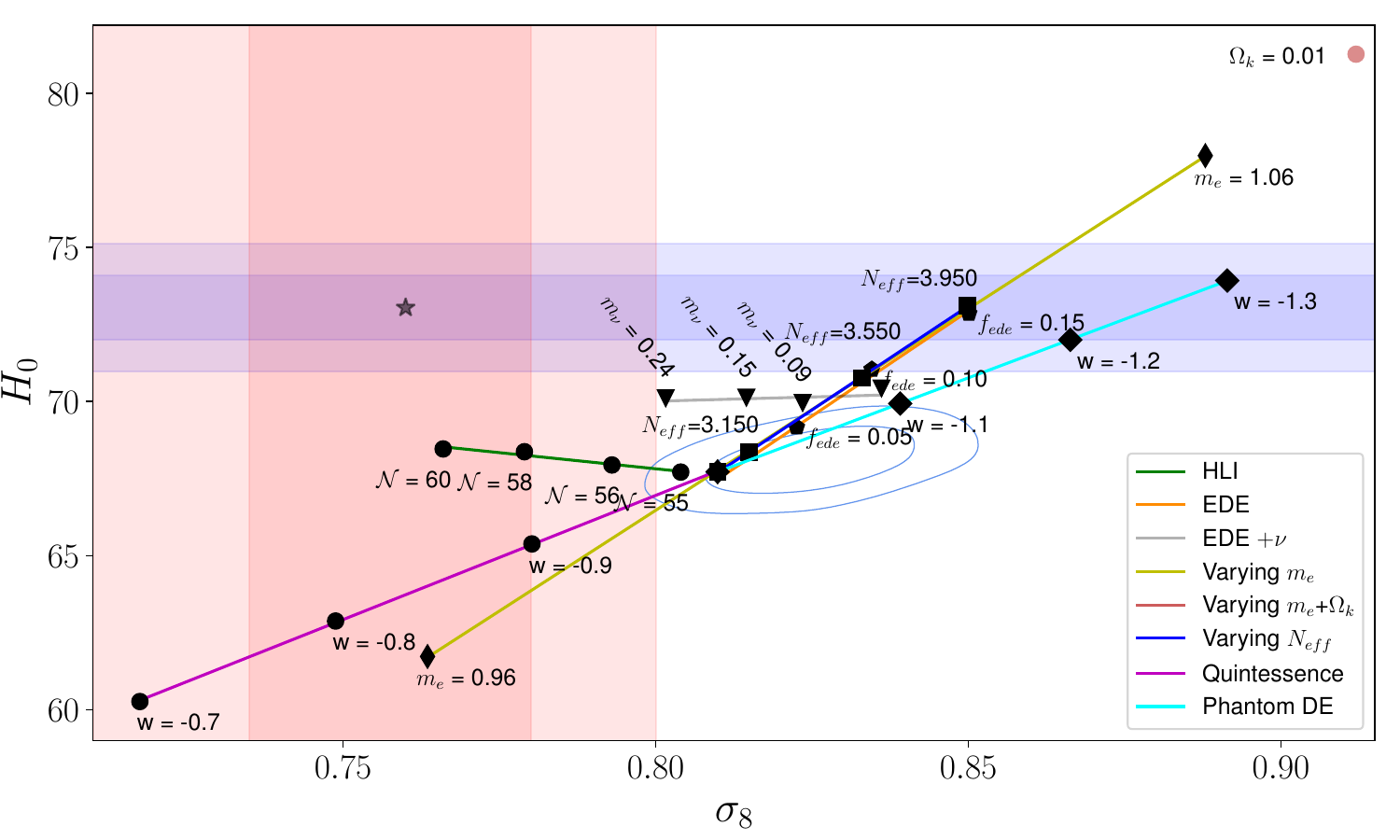}
  \caption{Relation between $H_{0}$ and $\sigma_{8}$ for different extensions of $\Lambda$CDM. The blue horizontal region represents the $1\sigma$ and $2\sigma$ contour from SH$0$ES~\citep{Riess:2021jrx} and the vertical pink region represents the $1\sigma$ and $2\sigma$ contour from KiDS-1000 3x2pt analysis~\citep{Heymans:2020gsg}. The star represents the mean value of the constraints from SH0ES and DES. In blue, we show the marginalized $68\%$ and $95\%$ confidence interval obtained by the \textit{Planck} collaboration \cite{Planck:2019nip}.}
  \label{fig:tensions}
  \centering
\end{figure*}

Figure \ref{fig:tensions} shows the constructed $H_0-\sigma_8$ plane for the proposed models. The numerical results can be seen in Appendix~\ref{sec:Appendix}. Figure~\ref{fig:tensions} allows us to readily visualize the tendency of the models to address the $H_0$ and $\sigma_8$ tensions. To help guide the eye, we add the $68\%$ and $95\%$ credible intervals of $H_0$ from SH$0$ES~\citep{Riess:2021jrx} as the blue shaded region, and of $\sigma_8$ from KiDS-1000 3x2pt analysis~\citep{Heymans:2020gsg} as the red shaded region. The blue contour shows the two-dimensional marginalized $68\%$ and $95\%$ 
credible intervals obtained by the \textit{Planck} collaboration \cite{Planck:2019nip} assuming the $\Lambda$CDM model.

Taking both the $H_0$ and $\sigma_8$ tensions at face value, models that could potentially address both tensions should have their empirical line passing through the overlap region between the blue and red shaded regions. Most of the models considered here have indeed a different direction. The EDE model, varying $N_\mathrm{eff}$ or $m_e$, and modified DE show a \textit{higher} $\sigma_8$ for higher values of $H_0$, i.e. a positive correlation between $\sigma_8$ and $H_0$. These models were invoked with the intention of solving the Hubble tension, however, the increase in $H_0$ comes at the cost of an increase in $\sigma_{8}$. 
The correlation between $H_0$ and $\sigma_8$ is already known for some of the models but is presented here -- to the best of our knowledge -- for the first time for other models, like the varying $m_e$ and HLI models. Another interesting detail that can be read from this plot is that for the varying-$m_e$ and quintessence models, there is a region in parameter space where these models can exhibit lower values of $\sigma_8$ than the \textit{Planck} value (assuming $\Lambda$CDM), i.e.\ relaxing the $\sigma_8$ tension. But this comes at the cost of lowering the value of $H_0$, i.e.\ worsening the $H_0$ tension.

Two models that we considered, the EDE+neutrinos and the HLI models, present a different correlation between $H_0$ and $\sigma_8$. 
In the case of EDE+neutrinos the value of $H_0$ is almost constant at a value around $H_0\sim 70$ km/s/Mpc while $\sigma_8$ decreases, relaxing the $\sigma_8$ tension. 
The HLI model, in turn, shows a negative correlation between $H_0$ and $\sigma_8$, i.e.\ as we vary $N_\mathrm{e-fold}$ the value of $H_0$ increases while $\sigma_8$ decreases. 
We will discuss the detailed behavior of the models below.

\vspace{0.5cm}


In order to understand the interplay between the new physics and the parameters $H_0$ and $\sigma_8$, one must remember that the $H_0$ tension can also be thought of as a discrepancy in the sound horizon,

\begin{equation}
 r_{s}(z_{\star})= \int^{\infty}_{z_\star} \frac{c_s(z)}{H(z)} dz, \,
\end{equation}
which should be reduced by $\approx 5\%$ in order to solve
the tension. Above, $c_s(z)$ is the sound speed of the photon-baryon fluid and $z_{\star}$ is the redshift of the last scattering. 
As it is well known, a change in the sound horizon can affect the angular size of the sound horizon according to the relation:
$\theta_{s}(z_{\star}) =r_{s}(z_{\star})/D_{A}(z_{\star})$, 
where 
\begin{equation}
  D_{A}(z_{\star})= \frac{1}{1+z_\star} \int_0^{z_\star} \frac{1}{H(z)} dz \, .
\end{equation}
The quantity $\theta_{s}(z_{\star})$ is very well measured from the
CMB power spectra (in particular the position of the first acoustic
peak). Therefore, any viable modification to the standard cosmological model, while changing the values of $r_{s}(z_{\star})$ and $D_{A}(z_{\star})$ (and consequently changing $H_0$), should
not alter the standard prediction for $\theta_{s}(z_{\star})$. 

In the following, we discuss the physics behind the observed $H_0$-$\sigma_8$ correlation model by model. 

\begin{itemize}

  \item HLI: The HLI model shows a negative correlation between $H_0$ and $\sigma_8$, i.e.\ as we vary $N_\mathrm{e-fold}$ the value of $H_0$ increases while $\sigma_8$ decreases. Moreover, when extrapolating the $H_0-\sigma_8$ line of the HLI model until the SH$0$ES $H_0$ value, the corresponding $\sigma_8$ value would be below the KiDS $\sigma_8$ value. While larger values of $N_\mathrm{e-fold}$ are allowed by theory, they are not enough to reach sufficiently high values of $H_0$ with a reasonable inflation duration. Nevertheless, the pattern of this model is very interesting because it sheds light on the role that a primordial term modification, such as inflationary theory, can play in the tension under analysis. Modulating the duration of inflation minimally impacts the current expansion rate of the Universe, as the primordial spectrum does not directly factor into the background equation. Instead, the spectrum of primordial scalar perturbations enters the $\sigma_8$ equation, as these are the seeds that determine the clustering of matter during the Universe's evolution. Now, the longer the duration of inflation, the greater the $k$ mode corresponding to the entry into the horizon at the time of the CMB, and thus the smaller the amplitude of the inflationary potential \cite{Mortonson:2010er}, since we assume a primordial spectrum with a spectral index lower than unity. This means that the greater $\mathcal{N}$ the lower the value of $\sigma_8$, i.e. the lower the amplitude aggregation of structures.

  \item EDE: The behavior of the EDE model with respect to the $H_0$ and $\sigma_8$ tensions has been well studied in the literature \cite{Vagnozzi:2021gjh, Ye:2021nej, Abdalla:2022yfr, Poulin:2023lkg}. The positive correlation between the two parameters can be understood as a consequence of preserving the model's fit to the CMB data, in particular, the amplitude of the early integrated Sachs-Wolfe (eISW) effect: Since EDE boosts the expansion rate before recombination, it leads to a suppression of the growth of gravitational potential wells. The growth of potential wells, however, is well constrained by the eISW effect, which itself can be obtained from the amplitude of the acoustic peaks, especially the first, of the CMB power spectrum. In order to preserve the fit to the CMB data, the matter density, $\Omega_\mathrm{m}$, is increased, which in turn leads to a higher amplitude of matter clustering, $\sigma_8$, at late times \cite{Vagnozzi:2021gjh}. Further, an increase of the expansion rate before recombination leads to increased damping on small scales of the CMB, which can be compensated by a higher $\Omega_\mathrm{m}$ and higher $n_s$ \cite{Ye:2021nej}, again resulting in a higher $\sigma_8$.
  
  \item EDE+$\boldsymbol{m_\nu}$: In this model, the value of $H_0$ is almost constant at a value around $H_0\sim 70$ km/s/Mpc while $\sigma_8$ decreases, relaxing the $\sigma_8$ tension. The approximately constant value of $H_0$ is a result of increasingly higher preferred values of $f_\mathrm{ede}$ for higher fixed values of $m_\nu$. As discussed above, a higher $f_\mathrm{ede}$ results in a higher $\sigma_8$, as a consequence of preserving the amplitude of the eISW effect and the small-scale damping. Massive neutrinos, on the other hand, suppress the growth of structure in the late Universe due to their noncold nature, which allows them to escape and smooth out potentials (free streaming), which in turn decreases $\sigma_8$~\cite{Lesgourgues:2006nd}. However, a higher $f_\mathrm{ede}$ and a higher $m_\nu$ come at the cost of a worsened goodness of fit to the data~\cite{Reeves:2022aoi}. The EDE+neutrino model does not have a $\Lambda$CDM limit, as opposed to the other models considered here, since the parameters of the EDE model $\{f_\mathrm{ede}, z_c,\theta_i\}$ are left free to vary during the analysis.

 \item $N_\mathrm{eff}$:  
  As one of the most well-known extensions of the standard cosmological model, the impact of varying $N_\mathrm{eff}$ in the cosmological tensions has been largely discussed in the literature \cite{Knox:2019rjx, Schoneberg:2021qvd,Archidiacono:2013fha}. 

  The effect of an increase in $N_\mathrm{eff}$ from the standard model value is very similar to the case of EDE or other models that lead to a decrease of the sound horizon at recombination, although the physics of adding a new relativistic species is different. A larger value of $N_\mathrm{eff}$ leads to a decrease in the sound horizon at recombination since this extra radiation increases the radiation energy density, increasing the expansion rate $H(z)$. Keeping $\theta_s$ fixed and decreasing $r_s$ results in a decrease in $D_A$ and a consequent increase in the late time expansion rate, $H_0$. Adding extra relativistic species also delays the time of the matter and radiation equivalence, and increases the damping scale, causing a similar decrease of $D_A$ at fixed damping angular scale $\theta_d = r_d/D_A$. These effects lead to a shift and smear of the acoustic peaks.
  Like in the EDE case, this leads to an enhancement in the amplitude of the eISW effect.
  Again, to preserve the fit to the CMB data, the matter density is increased, leading to a higher $\sigma_8$. 
  This positive correlation between $H_0$ and $\sigma_8$ in this model is what we see in Fig.~\ref{fig:tensions}.
  
  \item $\delta m_\mathrm{e}$ and $\delta m_\mathrm{e} + \Omega_k$:
  When altering the electron mass, we are affecting the ionization history during recombination, by changing the energy level of hydrogen ($E_{\mathrm{H}} \sim m_{e}$) and the Thomson scattering cross section ($\sigma_{\mathrm{T}} \sim 1/m_e^2$), affecting the recombination redshift. Therefore, if the electron mass was bigger in the early times compared to late times, the energy necessary to ionize the hydrogen atoms gets larger and the Thomson scattering cross section gets smaller. This leads to an earlier recombination, reducing the sound horizon and the Silk damping scale.
  This model has a positive correlation between the electron mass and the Hubble constant. In the same way as the models described above, this model also suffers from the \(H_0-\omega_m\) degeneracy, which raises the matter density while the current expansion rate grows, worsening the $\sigma_8$ tension. On the other hand, if one lowers the electron mass at early times, then the $\sigma_8$ value can be in agreement with the local measurements, despite worsening the $H_0$ tension.

  Adding a small amount of curvature in this model does not change the correlation of the parameters, but allows a higher value of $H_0$ and $\sigma_8$ to be obtained for the same electron mass when compared with the case with zero curvature.

\end{itemize}

\vspace{0.5cm}

As we saw above, most of the early time solutions to the Hubble tension that reduce the sound horizon lead to an increase in $\omega_m$, which worsens the $\sigma_8$ tension. That is the reason we see the positive correlation between $H_0$ and $\sigma_8$ in the plot for many of the early time solutions like the EDE, varying $m_e$ and varying $N_\mathrm{eff}$ models. It is difficult to decrease the sound horizon without changing $\omega_m$ and fully solve the tensions unless the pre-recombination solution has a mechanism to avoid that, like the EDE+$m_\nu$ model.

Another possibility is considering new physics post-recombination, known as the late time solutions to the Hubble tension. In these solutions, we change the late time evolution of the angular diameter distance, $D_A(z)$, 
for redshifts lower than $z_\star$,
without changing $D_A(z_\star)$ and $r_s(z_\star)$.

\begin{itemize}
  \item Modified DE: 
  One of the possible late time solutions is modifying the dark energy behavior with a constant dark energy equation state $w \neq -1$. By changing $w$, we modify the DE energy density, directly changing the late expansion rate of the Universe. To get an increase in $H_0$, we need the energy density of dark energy to be increasing with time. In this scenario, this is possible by considering phantom dark energy $w < -1$ ~\cite{DiValentino:2016hlg, Vagnozzi:2019ezj}. Since the dark energy density is increasing in order to obtain higher values of $H_0$, its energy density was smaller than what we predicted for $\Lambda$CDM in the past. Therefore, to maintain a flat Universe $\Omega_{\mathrm{total}} = 1$, a higher density of matter in the past is necessary, which in turn increases $\sigma_8$. 
  Having $w<-1$ also affects the growth of perturbations, both through its impact from the background expansion and the varied sound speed. This modified expansion of the Universe from the phantom DE leads to perturbations that cluster more than in the case of $\Lambda$CDM, which increases $\sigma_8$. This behavior can be seen in Fig.~\ref{fig:tensions} where we get a positive correlation between $H_0$ and $\sigma_8$ for the phantom DE model.
  
  For quintessence, with $-1 < w < -1/3$, we can only have smaller values of $H_0$ than the $\Lambda$CDM one, exacerbating the Hubble tension, as we can see in Fig.~\ref{fig:tensions}.
  Using only CMB, due to the geometrical degeneracy one cannot fully constrain $w$, while adding BAO or supernova data break this degeneracy allowing to constrain $H(z)$ \cite{Escamilla:2023oce}.

\end{itemize}
\vspace{0.5cm}

\begin{figure*} [t!]
\centering
\includegraphics[width=1.0\textwidth]{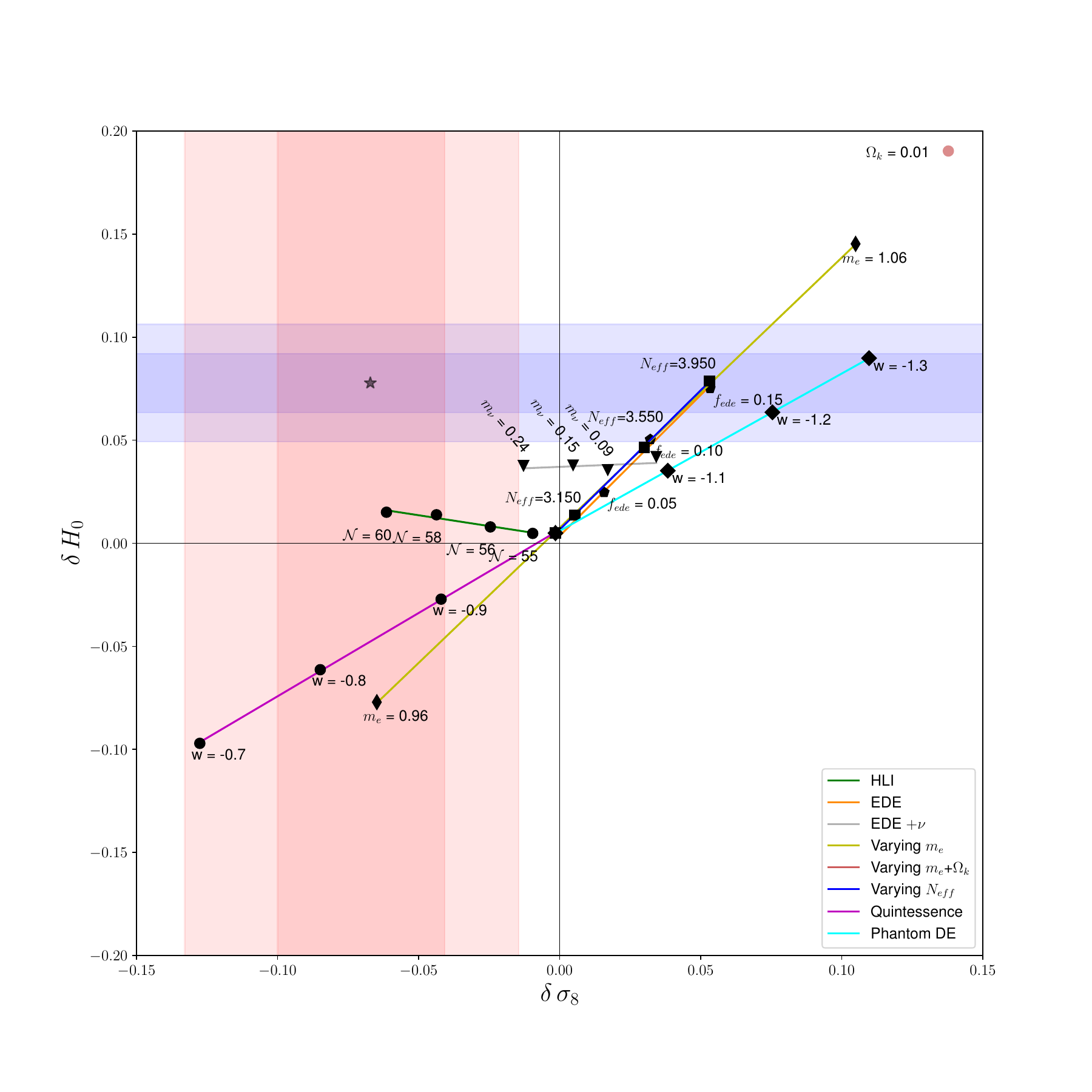}
\caption{Same as Fig.~\ref{fig:tensions} but showing the normalized quantities $\delta H_{0}$ and $\delta \sigma_{8}$ as defined in Eq.~\ref{eq:delta_H0_S8}.}
\label{fig:deltas}
\end{figure*}

An alternative to the $H_0-\sigma_8$ plane in Fig.~\ref{fig:tensions} is to plot the normalized quantities $\delta H_{0}$ and $\delta \sigma_{8}$, which are defined as
 \begin{equation}
   \delta H_0 \equiv \frac{{H_0}_\mathrm{(Ext)}-{H_0}_\mathrm{(CMB)}}{{H_0}_\mathrm{(Local)}}\,, \quad
  \delta \sigma_8 \equiv \frac{{\sigma_8}_\mathrm{(Ext)}-{\sigma_8}_\mathrm{(CMB)}}{{\sigma_8}_\mathrm{(Local)}}\,,
\label{eq:delta_H0_S8}
\end{equation}
where the quantities with subscript $\mathrm{(Ext)}$ denote the inferred parameters in the $\Lambda$CDM extension, ${H_0}_\mathrm{(CMB)}$ and ${\sigma_8}_\mathrm{(CMB)}$ are the mean values reported by the \textit{Planck} collaboration \cite{Planck:2018vyg}, ${H_0}_\mathrm{(Local)}$ is the mean value of the Hubble parameter reported by the SH$0$ES collaboration~\citep{Riess:2021jrx} and ${\sigma_8}_\mathrm{(Local)}$ the mean value of $\sigma_8$ reported by the KiDS-1000 collaboration for cosmic shear~\citep{Heymans:2020gsg}.
The $\delta H_0-\delta \sigma_8$ plane is shown in Fig.~\ref{fig:deltas}, and contains the same information as Fig.~\ref{fig:tensions} but has the advantage that one can compare the empirical lines with respect to the ``origin'', which represents the values preferred by the \textit{Planck} CMB data assuming the $\Lambda$CDM model. As we can see in Fig.~\ref{fig:deltas}, not all models have a $\Lambda$CDM limit and pass close to the origin. 

\subsection{Extensions of the visual tool}
\label{sec:chi2}

\begin{figure*} [t!]
  \centering
  \includegraphics[width=0.8\textwidth]{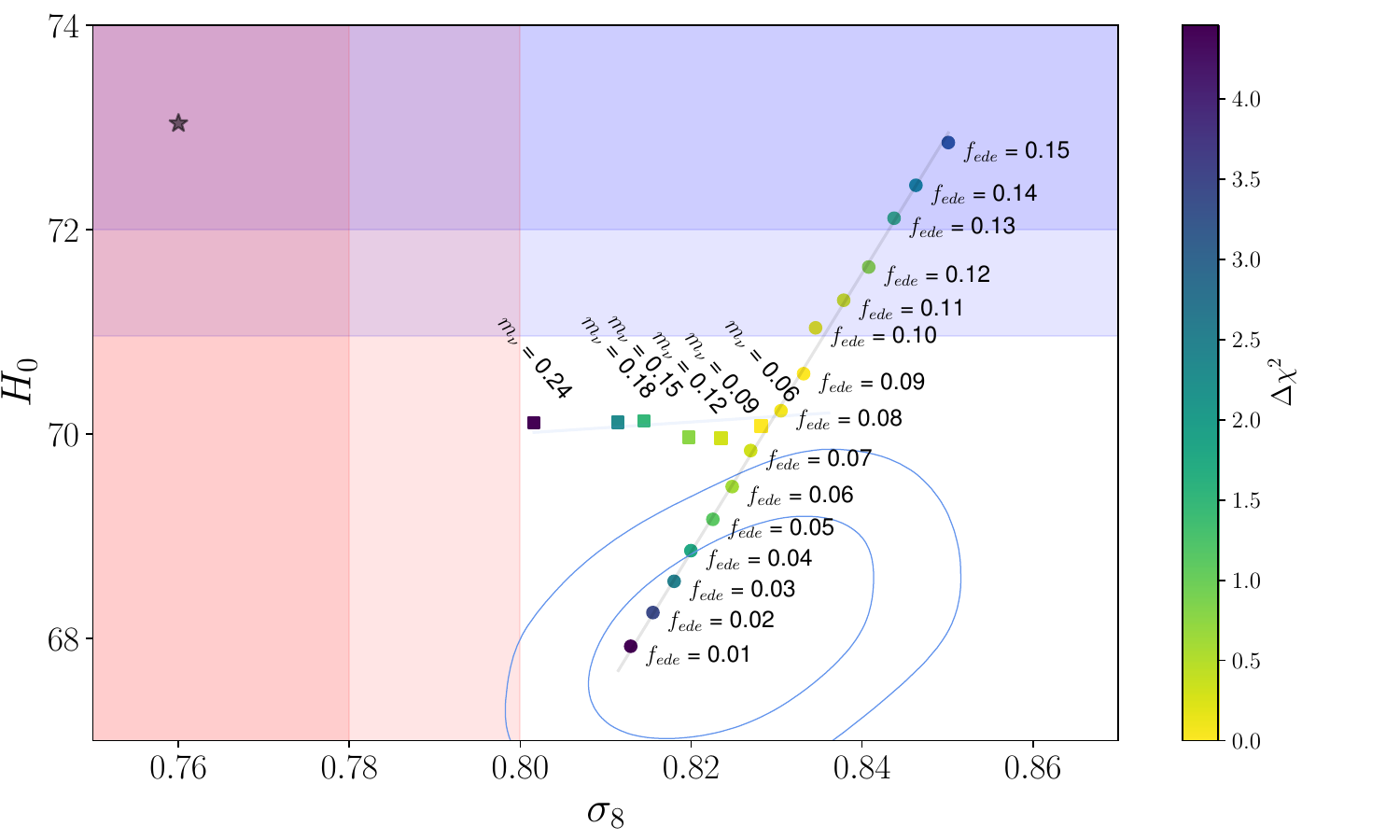}
  \caption{$H_{0}-\sigma_{8}$ plane for the EDE and EDE+$\nu$ models and the respective $\Delta \chi^2$ of each run. The blue horizontal region represents the 1-$\sigma$ and 2-$\sigma$ contour from SH$0$ES~\citep{Riess:2021jrx} and the vertical pink region represents the 1-$\sigma$ and 2-$\sigma$ contour from KiDS-1000 3x2pt analysis~\citep{Heymans:2020gsg}. In blue, we show the marginalized $68\%$ and $95\%$ confidence interval obtained by the \textit{Planck} collaboration \cite{Planck:2019nip}.}
  \label{fig:chi2}
\end{figure*}

The visual tool we presented in the previous section has the goal of illustrating the correlation between $H_0$ and $\sigma_8$, in order to show whether there is a region in the parameter space where the model's prediction can reach values of $H_0$ that are compatible with SH$0$ES and/or values of $\sigma_8$ that are compatible with KiDS. It does not tell us, however, whether the model can solve the tensions since it does not take into account the quality of the fit to the data. The values of the model parameters that solve the tensions might not correspond to the ones preferred by the data but might be ruled out by the data.

To visualize not only the relation between cosmological parameters but also the region of parameter space that is preferred by the data, we color code the $H_0-\sigma_8$ plane according to the goodness of fit, $\Delta \chi^2 (\theta) = -2 \ln (\mathcal{L} (\theta)/\mathcal{L}_\mathrm{max})$. The place where the $\chi^2$ is minimum corresponds to the best fit parameters. As an example, in Fig.~\ref{fig:chi2}, we show this for the $H_0$ and $\sigma_8$ values of the EDE and EDE+$\nu$ models obtained from a minimization in \citep{Herold:2021ksg, Reeves:2022aoi}. We can see that the minimum $\chi^2$ does not correspond to the highest values of $H_0$, which lie within the SH$0$ES 68$\%$ credible interval (blue-shaded region). In the case of the EDE+$\nu$ model, the empirical line may suggest that the model is able to resolve both tensions, but the region where it reaches the highest values of $H_0$ and smallest values of $\sigma_8$ are disfavored by the data.

Therefore, although the correlation between $H_0$ and $\sigma_8$ might indicate the possibility of a model to solve both tensions, it is important to consult other statistics, like
the goodness of fit to the data, in order to conclude whether a model can solve the cosmological tensions. Moreover, the method used above does not include the variance or error bar of the parameters, which needs to be taken into account for a complete analysis.


\section{Conclusions}
\label{sec:Conclusions}

In this paper, we present a visualization tool, the $H_0$-$\sigma_8$ plane, that provides an easy way to explore the correlation between the $H_0$ and $\sigma_8$ parameters for different cosmological models. 
This method is inspired in \citep{Vagnozzi:2019ezj}, from which we went further in elaborating a more general illustrative representation to show the tendency of broad classes of models to solve one or both cosmological tensions. Such a method maintains the advantages of its parent method, namely that fixing one or a few extra parameters of the model does not inflate the error bars compared to $\Lambda$CDM and, hence, leads to an easier visualization of the correlation between $H_0$ and $\sigma_8$ ~\cite{Vagnozzi:2019ezj,deSa:2022hsh}.
Since the analysis is done in a lower-dimensional parameter space, the method allows for a simple check of the tendency between $H_0$ and $\sigma_8$.
We expect that this visualization method is particularly useful for models that are affected by prior volume effects in the MCMC analysis, where this tool allows us to explicitly read the correlation between $H_0$ and $\sigma_8$, information that might be hidden in an MCMC posterior. 

We constructed the $H_0$-$\sigma_8$ plane for a few representative models that claim to address the tensions, choosing a heterogeneous sample of behaviors for a better illustration of the tool. One or more of the beyond-$\Lambda$CDM parameters of these models is fixed and a statistical analysis is performed. We performed an MCMC analysis for most models and a minimization for the models that are affected by prior volume effects. We showed that the resulting $H_0$-$\sigma_8$ plane can illustrate very clearly the correlation $H_0$ and $\sigma_8$ for these models. 
It is particularly interesting to apply this method to some of the most promising solutions to the $H_0$ tension, to understand their correlation with the $\sigma_8$ tension. Therefore, we analyze the varying electron mass and the EDE models, gold and silver medals in \citep{Schoneberg:2021qvd}. We can see that both present the same tendency: an increase in $H_0$ leads to an increase in $\sigma_8$, worsening the $\sigma_8$ tension. This tendency had already been pointed out in the literature for EDE~\citep{Hill:2020osr,Secco:2022kqg} but shown explicitly for the varying electron mass model. This tool also makes it easy to see the role of allowing the neutrino masses to be free in models like EDE, which changes the relation between $H_0$ and $\sigma_8$ compared to the model with fixed neutrino masses.
Another interesting case is the HLI model, which presents a negative correlation between $H_0$ and $\sigma_8$, i.e. the relation that is most promising for simultaneously solving both tensions. This is an interesting behavior of this model first reported here, which can be quickly visualized in the $H_0$-$\sigma_8$ plane.


Note that the simple $H_0$-$\sigma_8$ plane shown in Sec.~\ref{sec:result1} can be improved to provide additional information. For example, while in its vanilla form, it does not quantify the goodness of fit (or $\chi^2$), this can be added through the color code presented in Sec.~\ref{sec:chi2}. With this extra information, one can assess if the tension-resolving part of the parameter space corresponds to the one that is favored by the data. 
In addition, the slopes of each model in the plot are related to the specific dataset used. Moreover, one can compare the results with different datasets to see how the slopes of the curves vary, testing how the parameter space is sensitive to a specific dataset. 
Finally, one can also add error bars on the $H_0$-$\sigma_8$ plane, if desired. In general, this tool is very versatile and can be expanded as desired by adding additional information. 

In this work, we focused on the cosmological tensions and constructed the $H_0$-$\sigma_8$ plane, but this method could be used as a fast way of visualizing the correlation between any other cosmological parameters of interest. We envision the presented method could be useful as a quick illustrative tool that aids in model building, particularly for complex models with many parameters, which are not well constrained by observational data.

\section*{Acknowledgments}
I.O.C.P. is supported by the Fundacao Carlos Chagas Filho de Amparo a Pesquisa do Estado do Rio de Janeiro (FAPERJ), Grant No.
E-26/203.079/2023.
M.B. acknowledge Instituto Nazionale di Fisica Nucleare (INFN), sezione di Napoli, iniziativa specifica QGSKY. L.L.G is supported by the Fundacao Carlos Chagas Filho de Amparo a Pesquisa do Estado do Rio de Janeiro (FAPERJ), Grant No. E-26/201.297/2021, and is also supported by CNPq Grant nº. 307636/2023-2. L.L.G. would like to thank IPMU-Kavli Institute for the Physics and Mathematics of the Universe for warm hospitality during the period that part of this research was developed. 

We also acknowledge the use of CosmoMC and MontePython package. This work was developed thanks to the National Observatory (ON) computational support. This paper is based upon work from the COST Action CA21136, addressing observational tensions in cosmology with systematics and fundamental physics (CosmoVerse) supported by COST (European Cooperation in Science and Technology).


\appendix
\section{PARAMETER'S VALUES IN $\Lambda$CDM EXTENSIONS.}
\label{sec:Appendix}

Here we present the complete results that were used to construct the $H_0$-$\sigma_8$ plane in Sec.~\ref{sec:Results}. The values of $H_{0}$ and $\sigma_{8}$ considered in each extension of $\Lambda$CDM are shown in the tables below: 

\vspace{0.8cm}

{\bf \footnotesize{Nonminimal Higgs-Like Inflation (HLI)}}

\begin{table}[H]

\caption{\label{tab:HLI} We report here the constraints of the nonminimal Higgs-like inflation model obtained in Ref.~\cite{Rodrigues:2023kiz} by setting the $N_\mathrm{e-fold}=\mathcal{N}$ to fixed values. All the value are in $68\%$ C.L. , $H_{0}$ is in units of km/s/Mpc. }
\centering

\begin{tabular}{l|c|c}
\hline
& $H_0$ & $\sigma_8$ \\
\hline
$\mathcal{N}$=55 & $67.71 \pm 0.44$ & $0.804 \pm 0.003$ \\
$\mathcal{N}$=56 & $67.94 \pm 0.45$ & $0.793 \pm 0.003$ \\
$\mathcal{N}$=58 & $68.37 \pm 0.39$ & $0.779 \pm 0.004$ \\
$\mathcal{N}$=60 & $68.46 \pm 0.38$ & $0.766 \pm 0.005$ \\ 
\hline
\end{tabular}
\end{table}

\vspace{0.8cm}

{\bf \footnotesize{Early Dark Energy}}

\begin{table}[H]
  \centering
  \begin{tabular}{l|c|c|c}
    \hline
    EDE & $H_0$ & $\sigma_8$\\
    \hline
    \textbf{$f_{\mathrm{ede}} = 0$} & $67.60 $ & $0.811 $ & \\
    \textbf{$f_{\mathrm{ede}} = 0.05$} & $69.17 $ & $0.823$ &\\
    \textbf{$f_{\mathrm{ede}} = 0.10$} & $71.04 $ & $0.835$ &\\ 
    \textbf{$f_{\mathrm{ede}} = 0.15$} & $72.85 $ & $0.850 $ &\\
    \hline
    \hline
    EDE+$m_{\nu}$ & $H_0$ & $\sigma_8$ & $f_{\mathrm{ede}}$\\
    \hline
    $m_\nu = 0.06$ & $70.08$ & $0.8282$ & $0.0773$ \\
    $m_\nu = 0.09$ & $69.96$ & $0.8235$ & $0.0818$ \\
    $m_\nu = 0.15$ & $70.13$ & $0.8145$ & $0.0993$ \\
    $m_\nu = 0.24$ & $70.11$ & $0.8016$ & $0.117$ \\
    \hline
  \end{tabular}
  \caption{\label{tab:f_ede} best fit parameters in the early dark energy model, with standard and nonstandard neutrino mass. The first column shows the values of the fixed parameter of the theory, $f_{\mathrm{ede}}$ for the EDE model and $m_{\nu}$ for the EDE+neutrinos. The analysis for the EDE and EDE+$m_{\nu}$ models were performed in Refs. \cite{Herold:2021ksg} and \cite{Reeves:2022aoi}, respectively. $H_{0}$ is in units of km/s/Mpc.}
  \label{tab:EDE-EDE+nu}
\end{table}

\vspace{0.8cm}

{\bf \footnotesize{Extra Number of Effective Relativistic Species}}

\begin{table}[H]
\centering
\begin{tabular}{l|c|c}
\hline
& $H_0$ & $\sigma_8$\\
\hline
$N_{\mathrm{eff}}=3.046$  & $67.72 \pm 0.41$  & $0.8099 \pm 0.0059$  \\
$N_{\mathrm{eff}}=3.15$  & $68.36 \pm 0.42$  & $0.8149 \pm 0.0059$   \\
$N_{\mathrm{eff}}=3.55$  & $70.76 \pm 0.42$  & $0.8330 \pm 0.0061$ \\
$N_{\mathrm{eff}}=3.95$  & $73.11 \pm 0.46$  & $0.8499 \pm 0.0067$ \\
\hline
\end{tabular}
\caption{\label{tab:Neff} Parameter constraints for the $\Lambda$CDM+$N_{\mathrm{eff}}$ models. The first column shows the values of $N_{\mathrm{eff}}$ set for the extended model. All the values are in $68\%$ C.L. , $H_{0}$ is in units of km/s/Mpc.
A full analysis is presented in Ref.~\cite{deSa:2022hsh}.
}
\end{table}

\vspace{0.8cm}

\vspace{0.8cm}

{\bf \footnotesize{Varying Effective Electron Mass and Curvature}}

\begin{table}[H]
\centering
\begin{tabular}{l|c|c}
\hline
& $H_0$ & $\sigma_8$ \\
\hline
$ \delta m_e=0.96$ , $\Omega_k=0$ & $61.73 \pm 0.40$  & $0.7635 \pm 0.0065$ \\
$\delta m_e=1.00$ , $\Omega_k=0$ & $67.72 \pm 0.41$  & $0.8099 \pm 0.0059$  \\
$\delta m_e=1.06$ , $\Omega_k=0$ & $77.97 \pm 0.56$  & $0.8879 \pm 0.0079$ \\
\hline
\hline
$\delta m_e=1.06$ , $\Omega_k=0.01$ & $81.26 \pm 2.3$  & $ 0.9120 \pm 0.0081$ \\
\hline
\end{tabular}
\caption{\label{tab:me_me+k} Parameter constraints for the $\Lambda$CDM+$m_e$ and $\Lambda$CDM+$m_e$+$\Omega_k$ models. The first column shows the values of $m_e$ and $\Omega_k$ set in the analysis. All the value are in $68\%$ C.L. , $H_{0}$ is in units of km/s/Mpc.}
\end{table}

{\bf \footnotesize{Modified Dark Energy}}

\begin{table}[H]
\centering
\begin{tabular}{l|c|c}
\hline
& $H_0$ & $\sigma_8$ \\
\hline
$w=-0.7$  & $60.27 \pm 0.29$  & $0.7175 \pm 0.0070$   \\
$w=-0.8$  & $62.88 \pm 0.33$  & $0.7488 \pm 0.0064$  \\
$w=-0.9$  & $65.38 \pm 0.37$  & $0.7802 \pm 0.0062$  \\
$w=-1.0$  & $67.72 \pm 0.41$  & $0.8099 \pm 0.0059$  \\
$w=-1.1$  & $69.93 \pm 0.45$  & $0.8391 \pm 0.0060$   \\
$w=-1.2$  & $72.00 \pm 0.50$  & $0.8663 \pm 0.0063$  \\
$w=-1.3$  & $73.92 \pm 0.56$  & $0.8914 \pm 0.0066$  \\
\hline
\end{tabular}
\caption{\label{tab:w} Parameter constraints for the ${w}$CDM models. The first column shows the values of the equation of state, $w$, set in the analysis. All the value are in $68\%$ C.L., $H_{0}$ is in units of km/s/Mpc.
}
\end{table}


\bibliographystyle{apsrev4-1}
\bibliography{bibl}
\end{document}